\documentclass[prd,amsmath,amssymb,nofootinbib,twocolumn]{revtex4}
 \usepackage[english]{babel} 
 \usepackage{latexsym} 
 \usepackage{amsmath} 
 \usepackage{amsfonts} 
 \usepackage{amssymb} 
 \usepackage{textcomp} 
 \usepackage{hyperref}
 \usepackage{graphicx}  
 \usepackage{enumerate} 
 \usepackage{array} 
 \usepackage[latin3]{inputenc} 
 \usepackage{graphics} 
 \usepackage{color} 

 \newcommand{\be}{\begin{equation}} 
 \newcommand{\ee}{\end{equation}} 
 \newcommand{\ba}{\begin{eqnarray}} 
 \newcommand{\ea}{\end{eqnarray}}

\begin{document} 
\title{$p_t$-Angular power spectrum in ALICE events} 
\author{Felipe J. Llanes-Estrada and Jose L. Mu\~noz Martinez} 
\affiliation{ 
Departamento de F\'{\i}sica Te\'orica I, Universidad Complutense de Madrid, 
Parque de las Ciencias 1, 28040 Madrid, Spain.} 
\begin{abstract} 
We study the particles emitted in the fireball following a Relativistic Heavy Ion Collision with the  traditional angular analysis employed in cosmology and earth sciences, producing Mollweide plots of the $p_t$ distribution of a few actual, publically released ALICE-collaboration events and calculating their angular power spectrum. With the limited statistics at hand, we do not find evidence for acoustic peaks but a decrease of $C_l$ that is reminiscent of viscous attenuation, but subject to a strong effect from the rapidity acceptance which probably dominates (so we also subtract the $m=0$ component).
As an exercise, we still extract a characteristic Silk damping length (proportional to the square root of the viscosity over entropy density ratio). The absence of acoustic-like peaks is also compatible with a crossover from the QGP to the hadron gas  (because a surface tension at domain boundaries would effect a restoring force that could have driven acoustic oscillations).
Presently an unexpected depression appears in the l=6 multipole strength, which should be revisited by the ALICE collaboration with full statistics to confirm or discard it. 
\end{abstract}

\maketitle 

\section{Introduction}

The phase diagram of Quantum Chromodynamics is one of the guiding goals of much of the worldwide nuclear-particle physics efforts. Currently, the picture in which, at low baryon-density, the Quark-Gluon-Plasma (QGP) cools into a hadron medium through a smooth crossover finds wide support fundamented in lattice gauge theory computations~\cite{Fodor:2004nz,Chen:2004tb,Sharpe:2004ps}.
Additionally, studies of damping in the plasma and later hadron phase have been vigorously pursued~\cite{Davesne:1995ms,Dobado:2003wr,Csernai:2006zz}.

Empirical evidence for the crossover, other features of the phase diagram, or transport coefficients are less straightforwardly obtained, since experimental data show the conditions at the freeze out surface\footnote{It is generally believed that the freeze out happens after the system spends some time in a hadronic phase, but certain numerical fits~\cite{Rafelski:2015hta} suggest that there is little or no final state rescattering among hadrons after the collision,  meaning that chemical freeze out would occur right upon exiting the QGP.}.
For example, there is raging discussion on whether the critical end point of a first order phase transition present at larger baryon chemical potential has or has not been located~\cite{Lacey:2015yxg,Antoniou:2016hbx}.

It is therefore very reassuring when actual empirical evidence in support of the supposed phase diagram accrues, particularly the crossover at small baryon density, as for example the scaling of moments of the distribution for baryon-number fluctuations with volume (or as proxy, number of participants)~\cite{Mohanty:2009vb,Gupta:2009mu}. And we have relatively solid evidence that the ratio of viscosity to entropy density is low~\cite{ALICE:2016kpq,Gavin:2007zz} and not too far above the renowned $1/(4\pi)$ bound~\cite{Son:2007vk}.

Part of this contribution, based on a number of events publically released by the ALICE collaboration, is to observe that the absence of acoustic peaks in the angular spectrum of $p_t$ and related fluctuations (in addition to a clear effect of the rapidity acceptance cut) might be an additional hint of that crossover nature, and perhaps also attest attenuation (analogous to Silk damping in cosmology). 
This work follows on the footsteps of several others~\cite{Basu:2014ipa,Rafelski:2013obw,Dobado:2015vaa,Kuznetsova:2010pi} that exploit the
 similarity between the primeval cosmological explosion and Relativistic Heavy Ion Collision Experiments (RHIC-E).
We continue developing analysis methods, and, especially, try to apply them to study ALICE data in the public domain, setting the stage for studies with higher statistics.

The main analysis tool for the Cosmic Microwave Background (CMB) radiation is the angular spectral analysis used to describe temperature fluctuations in the sky at different angular resolution.
The spherical harmonic transform is an appropriate analysis method as the radiation is distributed over the celestial sphere. 
It has not escaped the attention of the RHIC-E community that the same method can be applied to radiation coming from {\it within} a sphere as opposed to entering it; some studies have provided Mollweide plots~\cite{Naselsky:2012nw} of particle distributions or angular spectra generated by Monte Carlo simulations.

To discuss RHIC-E, we need an appropriate coordinate system that matches the one used in cosmology. A natural one orients the $OZ$ axis along the beamline, and the polar angle $\theta$ is measured therefrom;  this can be traded for the rapidity $y\equiv\tanh^{-1}(v_{\parallel})=\tanh^{-1}\left(\frac{p_{\parallel}}{E}\right)$.
Particles (90$\%$ of them pions) have coordinates $(p_\parallel,{\bf p}_t)$ or equivalently,
$(p_t,y,\phi)$. Most often, rapidity is approximated by pseudorapidity $\eta\equiv-\ln{\left(\tan\frac{\theta}{2}\right)}$ that is more directly read from the pion track in such instruments as the time projection chamber of the ALICE experiment.

The temperature of the CMB is read from a black-body fit to photon wavelengths. In ALICE, there are relatively few particles (a few thousand in central collisions) so that the temperature map is subject to stronger statistical fluctuations.  To characterize mean temperature one may employ mean tranverse momentum~\cite{Torres-Rincon:2012sda}, with $E_t^2\equiv m^2+p_t^2$ because of its equilibrium distribution\footnote{Note that Eq.~(\ref{spectrum}) applies in the fluid's rest frame, but because of the expanding system, the pions finally detected are blue shifted~\cite{Stephanov:1999zu} according to $p_t\simeq p_t^{\rm rest}\sqrt{(1+\beta)/(1-\beta)}$; as this complicates the analysis significantly, we will not attempt to track the temperature back and remain in $p_t$ for the entire article.}
\begin{equation}\label{spectrum}
\frac{dN_i}{dy_p E_t dE_t}\sim\sqrt{T E_t}e^{-E_t/T}\ .
\end{equation}
The fluctuations of $p_t$ are on the other hand known to differ from temperature fluctuations (though divergences near the critical end point, for example, are expected to occur simultaneously in both)~\cite{Stephanov:1999zu}. But since $p_t$ is one of the more immediate observables,
we will map it out in RHIC-E as a surrogate of the temperature of the CMB, and use only averages thereof over the several particles emitted with approximately the same solid angle.

The temperature fluctuations of the CMB are usually divided in three pieces~\cite{munkhanov} 
\begin{equation}\label{3T}
\frac{\Delta T}{T}(\theta,\phi)=
 \left.\frac{\Delta T}{T}\right|_{\text{Doppler}}
 +\left.\frac{\Delta T}{T}\right|_{\text{SW}}
 +\left.\frac{\Delta T}{T}\right|_{\text{SWI}}
\end{equation}
where the Sachs-Wolfe piece stems from red/blue shifts of photons due to fluctuations in the gravitational potential (and density) at the time of last scattering; the integrated Sachs-Wolfe piece has the same physical basis but accumulates while the photons travel towards us; and the Doppler shift is the usual kinematic effect due to our motion (and gives a large dipolar radiation figure that masks any original dipole-like fluctuation). 

Likewise, we can divide the $p_t$ fluctuations in heavy ion collisions according to their physical origin,
\begin{equation}
\frac{\Delta p_t}{p_t}(\theta,\phi)  = \left. \frac{\Delta p_t}{p_t}\right|_{\text{jet}}
+ \left. \frac{\Delta p_t}{p_t}\right|_{\text{init}}
+ \left. \frac{\Delta p_t}{p_t}\right|_{\text{flow}}
+\left. \frac{\Delta p_t}{p_t}\right|_{\text{T}}\ . 
\end{equation}
We have separated the thermal fluctuations around equilibrium at freeze out (denoted with a $T$ subindex), those caused by collective flow, those coming from the initial nuclear states (perhaps a color glass condensate) and finally, the very strong ones caused by hard initial collisions (jets)\footnote{Sometimes the breakdown of the exponential in Eq.~(\ref{spectrum}), due to a slower power-law falloff appearing for high $p_t$, is adscribed to Tsallis statistics; we rather observe that hard QCD collisions are naturally power-law shaped as per the Brodsky-Farrar counting rules~\cite{Brodsky:1974vy} or Feynman-diagram natural dimension in jet production, so that if one is interested in statistical or fluid fluctuations, a cut excluding high-$p_t$ is in order, and Eq.~(\ref{spectrum}) is perfectly fine.}.
Indeed, jets should dominate the dipolar $C_1$ just as the kinematic Doppler effect dominates its
counterpart in the CMB. $C_1$ can be reduced with a $p_t$ cut, eliminating the hardest particles that come largely from jets.

\section{Multipole analysis}

As a function over the sphere, $\frac{\Delta p_t}{p_t}(\theta,\phi)$ may be expanded in orthonormal spherical harmonics, with coefficients
\begin{equation}\label{alminv}
a_{lm}=\int d\Omega\ Y^*_{lm}(\theta,\phi)\frac{\Delta p_t}{p_t}(\theta,\phi) \ .
\end{equation}

The angular (power) spectrum (of wide use in cosmology and earth sciences) is then 
\begin{equation}
\tilde{C}_l=\frac{1}{2l+1}\sum_m\left|a_{lm}\right|^2 \ .
\end{equation}
Obviously the phase information of the $a_{lm}$ is lost (several of the coefficients that we will report later in the ALICE context happen to be negative), and what remains is the angular spectral strength.
We use the notation $\tilde{C}_l$ to refer to this quantity when computed for only one collision event, and $\langle \tilde{C}_l\rangle$ to its ensemble average over a large number of collisions,
which should in the limit of an infinitely large ensemble, converge to $C_l$, 
\ba\label{Tcl}
\left\langle\left|\frac{\Delta p_t}{p_t}(\theta,\phi)\right|^2\right\rangle
&=&\sum_l \left\langle  \tilde{C}_l\right\rangle \sum_m\left|Y_{lm}(\theta,\phi)\right|^2 \\ \nonumber
&=&\sum_l\frac{2l+1}{4\pi}C_l \ .
\ea
In cosmology, the average is taken over the sky but not over multiple realizations of the system (as only one observable universe is at hand). 
An advantage in the field of Heavy Ion Collisions is that, as experimental repetition is no problem, the variance can be read off the data.
In cosmology, instead, its ``cosmic variance'' is 
\begin{equation}
\sigma_l\equiv\sqrt{\frac{\left\langle\left(\tilde{C}_l-C_l\right)^2\right\rangle}{C_l^2}}=\sqrt{\frac{2}{2l+1}}\ .
\end{equation}
Thus, while the estimator $\left\langle \tilde{C}_l \right\rangle$ in ALICE may be arbitrarily close to the true value $C_l$, there is a limit to the achievable precision in cosmology given by $\sigma_l$. This is because high multipoles can be studied  with different regions of the sky, but  low multipoles basically require integrals over the entire database, and thus only one measurement is possible.

The multipole index $l$ runs from $0$ to $l_{\rm max}$ given by the maximum angular resolution,
$l_{\rm max} \simeq \pi/\theta_{\rm res}$. In RHIC-E, this is limited by multiplicity: the several thousand particles in the most central collisions have to group in $(\theta,\phi)$ patches such that each of them may be considered a small continuum unit. This gives a coarse graining of the sphere surface that limits the reach in $l$ (so that the resulting collision-event maps look more like old COBE sky-maps rathen than the very detailed Planck ones, see later figure~\ref{fig:singleevents}).

\begin{figure}[hbtp]
\centering
\includegraphics[width=\columnwidth]{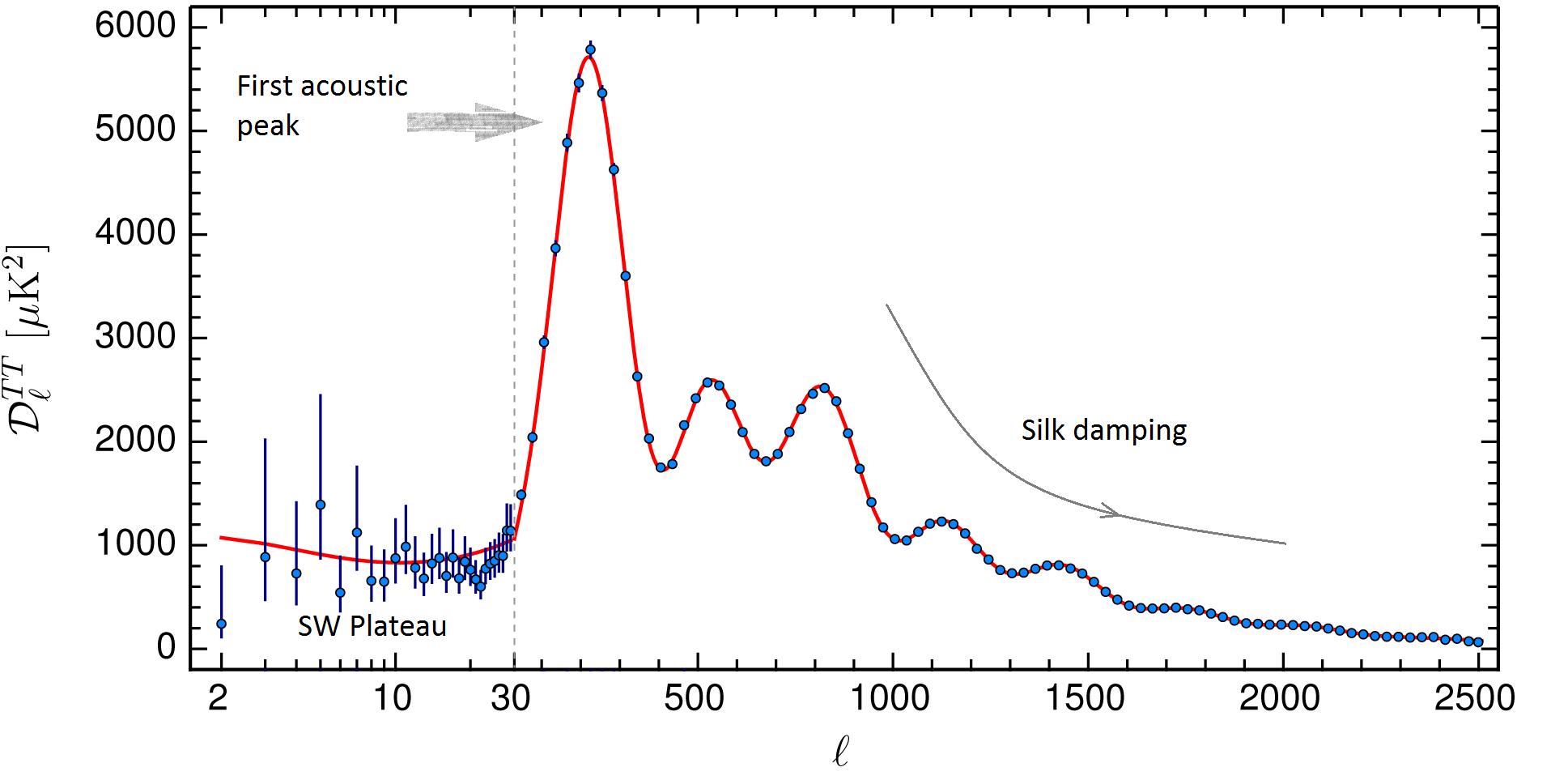}
\caption{Angular power spectrum of CMB temperature (Planck collaboration~\cite{planckinf}). 
The $y$ variable is $D_l^{TT}\equiv l(l+1)C_l/(2\pi)$; the solid line (red online) represents the standard cosmological ($\Lambda$CDM) model and the bars (blue online) the cosmic variance.
For the convenience of nuclear and particle physicists we have lettered a few features as described in the text.}
\label{specTTwmap}
\end{figure}

The resulting $ \tilde{C}_l$ angular spectrum observed in cosmology is reproduced in figure~\ref{specTTwmap}. 
The most salient features are the Sachs-Wolfe plateau for small $l<l_H$ (superHubble modes, those that were outside the horizon at decoupling time because of their large wavelength), followed by the acoustic peaks (subHubble modes) increasingly damped at large $l$.

\begin{figure}[hbtp]
\centering
\includegraphics[width=\columnwidth]{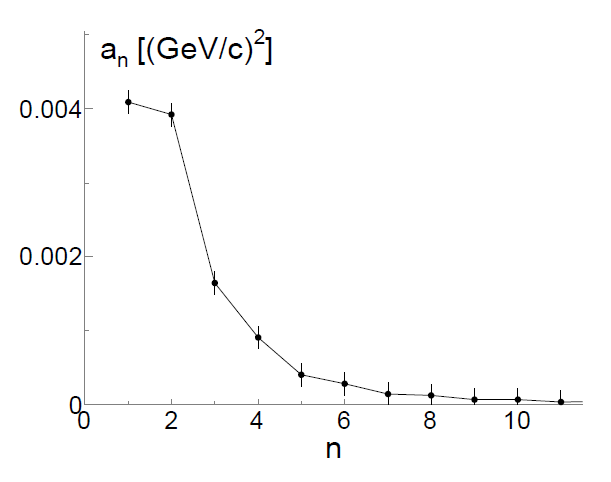}
\caption{Angular power spectrum calculated from two-particle $p_t$ correlations (from  STAR data) by M\'ocsy and Sorensen. Adapted from~\cite{Mocsy:2010um}; figure courtesy of the author.}
\label{anpwspct}
\end{figure}

Such acoustic oscillations stem from the dynamics of a repulsive force (the system pressure) and a restoring force (the gravitational attraction at inhomogeneities or potential wells). In conventional RHIC-E theory there is no such additional restoring force, with the effect of the strong interactions already accounted for in the equation of state (the pressure), so that we do not expect to see such peaks in a $p_t$ spectrum. Indeed, an angular spectrum produced by M\'ocsy and Sorensen~\cite{Mocsy:2010um} for two-particle $p_t$ correlations 
\begin{equation}
\left\langle
\frac{\Delta p_t}{p_t}(\theta,\phi)\frac{\Delta p_t}{p_t}(\theta' ,\phi' )
\right\rangle
=\frac{1}{4\pi}\sum_l(2l+1)C^{2p}_lP_l(\cos\Delta\theta)
\end{equation}
is replotted in figure~\ref{anpwspct} and shows no such peak, but one can argue that the maximum $l$ reach is very small. We will revisit the issue later in section~\ref{sec:interpret}.

As for the Sachs-Wolfe plateau, the angular spectrum of primordial temperature fluctuations
for those superHubble modes can be related to the scalar metric perturbations $P_R(k)$ at decoupling,
\begin{equation}\label{primorCl}
C_l=\frac{4\pi}{25}\int_0^{\infty}\frac{dk}{k}P_R(k)j_l^2(k\eta_0)\ ,
\end{equation}
expression known as ``primordial temperature fluctuation spectrum''.
These fluctuations accept, in standard cosmology, a power-law expression
\begin{equation}\label{PRAn}
P_R(k)=A_s^2\left(\frac{k}{k_0}\right)^{n_s-1}
\end{equation}\\
where  $A_s\sim 5\times 10^{-5}$ is the amplitude of the perturbations, $k_0=0.002-0.05\text{ Mpc}^{-1}$ an arbitrary, convention-dependent pivot scale, and $n_s$ the ``spectral index''.
The simplest models of inflation predict  $n_s\simeq1$ (confirmed by cosmological data that yield $n_s=0.97$). Taking it to be unity at face value, substituting  (\ref{PRAn}) in (\ref{primorCl}), and integrating, yields
\begin{equation} \label{powerClcosmo}
C_l=\frac{2\pi}{25}A_s\frac{\Gamma(3/2)\Gamma(1)\Gamma(l)}{\Gamma(3/2)\Gamma(l+2)}=\frac{2\pi}{25}A_s\frac{(l-1)!}{(l+1)!}=\frac{2\pi}{25}\frac{A_s^2}{l(l+1)} \ ,
\end{equation}
result valid for  $l<l_H$, and that provides a way to measure primordial fluctuations of the gravitational metric field ($A_s$) from temperature fluctuations today ($C_l$).
The denominator in Eq.~(\ref{powerClcosmo}) is what suggests to plot $l(l+1)C_l$ in figure~\ref{specTTwmap} because it is predicted to be $l$-independent (consistently with the CMB data, flat within the cosmic variance). But since this will not generally be the case in RHIC-E, we will multiply $C_l$ by different powers (or none) than in cosmology.

The $C_l$ coefficients can also be defined for many quantities of interest in RHIC-E, and it would be interesting to develop models linking them to properties of the hot nuclear medium.

To compute angular spectra from arbitrary data distributed over the sphere we employ the public software package {\tt SHTOOLS}~\cite{shtools}. From the suite of functions we highlight {\tt SHGLQ}, to compute Gauss-Legendre zeroes and weights over the sphere for integrations such as Eq.~(\ref{alminv}), and the less precise {\tt SHExpandLSQ} that, by least--squares optimization, can use arbitrary grids.

Another very common visualization tool for functions with domain on the sphere $S^3$ is the Mollweide projection. This maps a sphere of radius $R$ to the inside of an elipse in the $XY$ plane by means of
\ba\label{mollweq1}
x=\frac{2\sqrt{2}}{\pi}R\lambda\cos\beta \\
\label{mollweq2}
y=\sqrt{2}R\sin\beta\\
\label{mollweq3}
2\beta+\sin2\beta=\pi\sin\psi
\ea
where $\psi$ and $\lambda$ are, respectively, the latitude and longitude of a point over the sphere ($\psi=\pi/2-\theta$ and $\lambda=\pi-\phi$). Setting $R=1$ maps the unit sphere, and the intensity of the function of $(\theta,\phi)$ will be made visible by the color intensity of the point in that position. 

\section{Effect of the ALICE TPC and ITS acceptance}\label{sec:acceptance}

The major current detector for the study of Heavy Ion Collisions is ALICE at the LHC.
The instrument dedicated to measuring particle $p_t$ is the Time Projection Chamber (TPC)
which is supplemented by the Internal Tracking System for low-momentum hits (below about 200 MeV).
The ITS also assists the TPC even for faster particles (that the ITS cannot identify, but whose hits help projecting the traces to the collision point).

The TPC is not a hermetic detector but rather a central barrel: Its rapidity coverage is limited 
to about one unit of rapidity around the plane perpendicular to the beam axis. 

\begin{figure}[hbtp]
\centering
\includegraphics[width=\columnwidth]{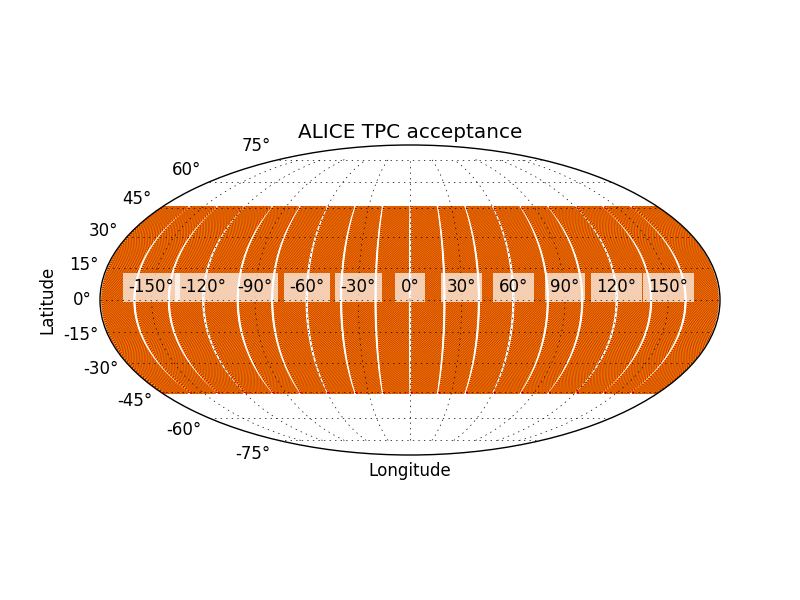}
\caption{Mollweide projection of a schematic acceptance function (Eq.~\ref{acepteq}) of the ALICE detector. The shaded area (orange online) represents the area where particle $p_t$ is measurable. Excluded are the polar caps (around the accelerated beams, where the nonhermetic TPC has no coverage) and support bars at 20$^o$ intervals.
\label{acceptancesim}}
\end{figure}

Additionally, there are support bars every 20$^o$ in azimuth, spanning 2$^o$, that detract from the instrumented region. 
therefore, we can define a TPC acceptance function
\begin{equation}\label{acepteq}
A(\theta,\phi)=\begin{cases}0\  \text{     if  }\theta\notin[44.25^o,135.7^o]\ \ \ or\ \ \  \phi\in\{20^o n\pm1^o\}\\1\  \text{     otherwise}\end{cases}
\end{equation}
that is what is actually represented in figure~\ref{acceptancesim}.

\begin{figure}[hbtp]
\centering
\includegraphics[width=\columnwidth]{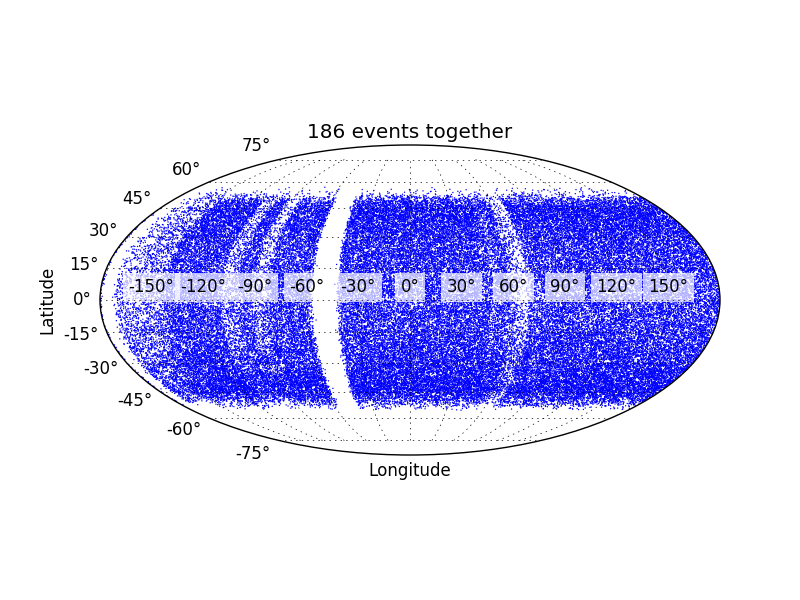}
\caption{186 ALICE events~\cite{cernopendata} in Mollweide projection; each point represents a charged particle at the shown angular position without regard to its $p_t$.}
\label{datvic}
\end{figure}

To see how this theoretical acceptance looks with real data, we have plotted 
186 events from the Pb-Pb ALICE 2.7 TeV data at various centralities~\cite{cernopendata}  in figure~\ref{datvic}. We ignore the statistical error associated with this data, so none will be quoted.

Each point in the figure corresponds to one particle produced in any of the events at the given angular position. From the plot we discern that the actual polar angle coverage extends a bit beyond the 44 degrees given in Eq.~(\ref{acepteq}), possibly due to the additional coverage provided by the Internal Tracking System (ITS) and other instruments. 

Additionally, not all the supposed azimuthal blanks from the TPC bars are there, though some seem visible.  there is an unexpected 20$^o$ hole in azimuthal coverage near -45$^o$ longitude that is probably due to instrumental failure\footnote{V. Gonz\'alez, in private communication, suggests that this reflects a blind spot of the innermost layers of ALICE's ITS. A different processing of the data that does not require a particle hit in those layers to accept a track might not show the void. Data analyzed without the inner layers is, to our knowledge, yet to be publically released}. 
The actual azimuthal acceptance shown in figure~\ref{datvic} seems therefore dominated by the Internal Tracking System and has a more complicated structure than the nominal TPC acceptance

\begin{figure}[hbtp]
\centering
\includegraphics[width=\columnwidth]{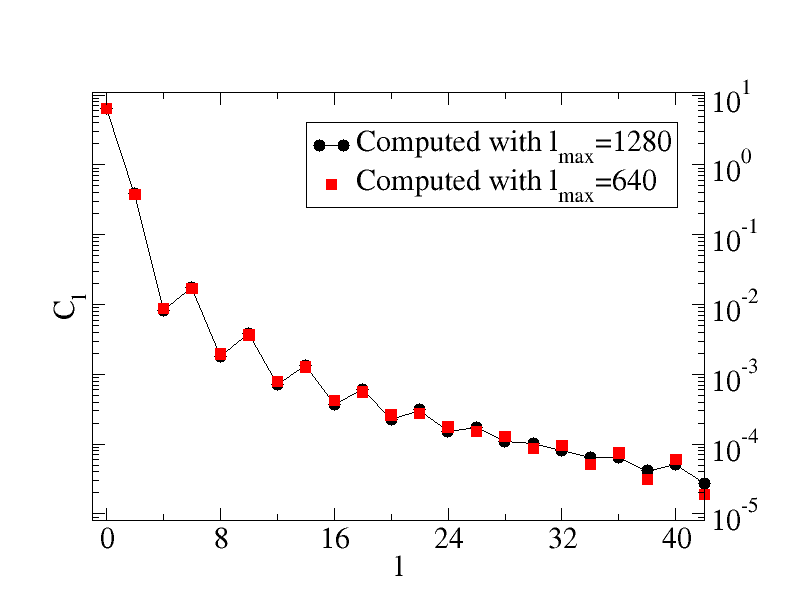}
\caption{Spherical harmonic transform (angular power spectrum) of ALICE's pseudorapidity acceptance function. Reflection symmetry around the equatorial plane (plane perpendicular to the accelerator beams) forces odd $l$ coefficients to vanish so we only represent even $l$.
}
\label{onlylatcut}
\end{figure}

We now show the $C_l$ angular spectrum resulting from the acceptance cuts alone (that is, for a particle distribution that is otherwise completely flat and structureless). 
We show in figure~\ref{onlylatcut} two computations of $C_l$ that consider only the polar (rapidity) cut in Eq.~(\ref{acepteq}) with $l_{\rm max}=640$ and $l_{\rm max}=1280$ to check for expansion and integration convergence, although only much smaller $l$s are relevant for few-thousand particle multiplicities. Note the staggering of the even $l$ $C_l$s.
It is worth remarking that, with our normalizations, the monopole $C_0$ coefficient of the unit function (=1 over the entire $\theta$, $\phi$ coverage) would be\footnote{Given a function $f(\theta,\phi)$, and its angular average $\bar{f}(\theta,\phi)\equiv \int d\Omega\ f(\theta,\phi)/4\pi$, the monopole $a_{00}$ coefficient (that yields $C_0$ upon squaring) can be computed as
$ a_{00}=\int d\Omega Y_{00}^*(\theta,\phi) (f/\bar{f})(\theta,\phi)=\frac{1}{\sqrt{4\pi}}\frac{\int d\Omega\ f(\theta,\phi)}{\frac{\int d\Omega\ f(\theta,\phi)}{4\pi}}=\sqrt{4\pi}
$.
}
 $4\pi\simeq 12.56$, so that the actual $C_0$ in figure~\ref{onlylatcut} is basically this number multiplied by the fraction of the rapidity-accepted solid angle.
Because the polar angle interval is $180^o$ and the gap in rapidity coverage at each of the poles is $45^o$, which is a fourth thereof, it is not surprising to see the data in figure~\ref{onlylatcut} staggering between $l=4n$ and $l=4n+2$ (odd multipoles vanish and are not shown at all).

\begin{figure}[hbtp]
\centering
\includegraphics[width=\columnwidth]{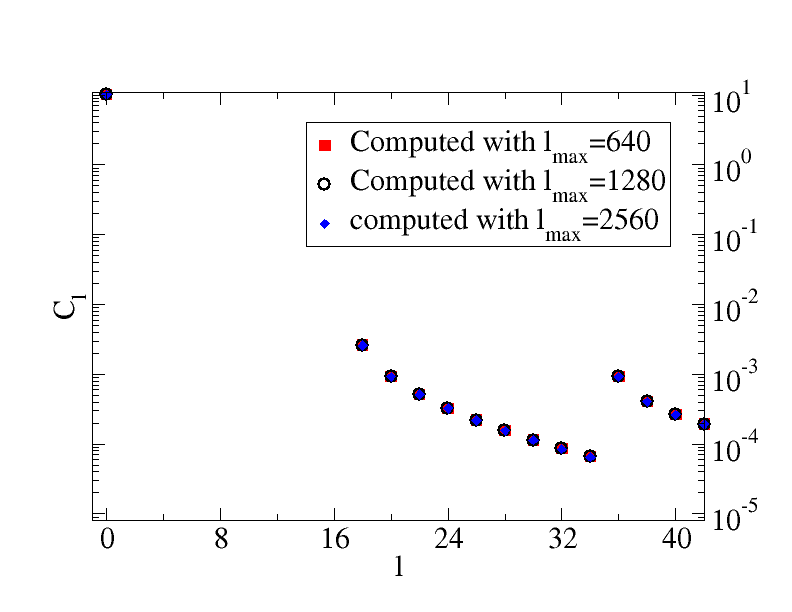}
\caption{Angular spectrum corresponding to a unit function $f=1$ over the entire sphere
except a set of $2^o$ broad longitude intervals around each meridian multiple of
 $20^o$, corresponding to the spacing between the bars of ALICE's TPC.
}\label{allbars}
\end{figure}

A similar calculation is shown in figure~\ref{allbars} for the azimuthal acceptance (ignoring the rapidity/polar angle cut in Eq.~(\ref{acepteq})\ ).
The coefficients $C_l$ with $l=2\dots 16$ as well as all the odd ones vanish, given the symmetry and 20$^o$ interval between bars, so that $C_0$ is first followed by $C_{18}$ with pattern repetitions every $18$ units of $l$ (overtones). 

\begin{figure}[hbtp]
\centering
\includegraphics[width=\columnwidth]{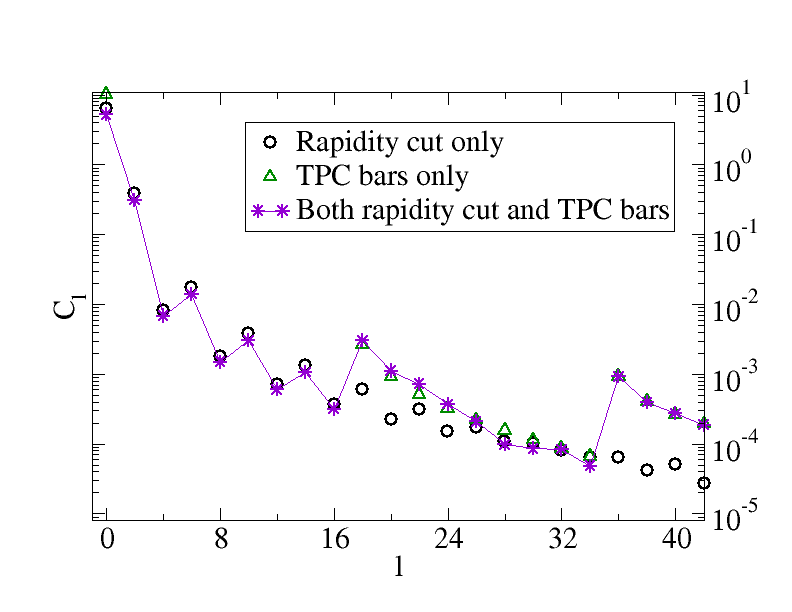}
\caption{Theoretical angular spectrum produced by the (simplified) TPC acceptance of ALICE's experiment.
\label{latandbars}}
\end{figure}

Both azimuthal and rapidity acceptance are plot together in figure~\ref{latandbars} which is the $C_l$ spectrum corresponding to Eq.~(\ref{acepteq}).

We easily see the maxima at $l$ multiples of 18 (due to the TPC bars).
Likewise, the multipoles 2 to 16 stem from the reduced pseudorapidity acceptance of the TPC.

As a last example of angular spectrum of the acceptance, we will consider the rapidity acceptance of the ITS. This is suggested by figure~\ref{datvic} where we see that the rapidity coverage of the data released extends well beyond the TPC acceptance cut.
According to its technical design report~\cite{ITS}, the ITS covers rapidity in $(-1.95,1.95)$, or in latitude,  $(-73.8^o,73.8^o)$, rather more extensive than the TPC, which may explain many of the higher latitude hits in the Mollweide projection of figure~\ref{datvic}. 

\begin{figure}
\includegraphics[width=\columnwidth]{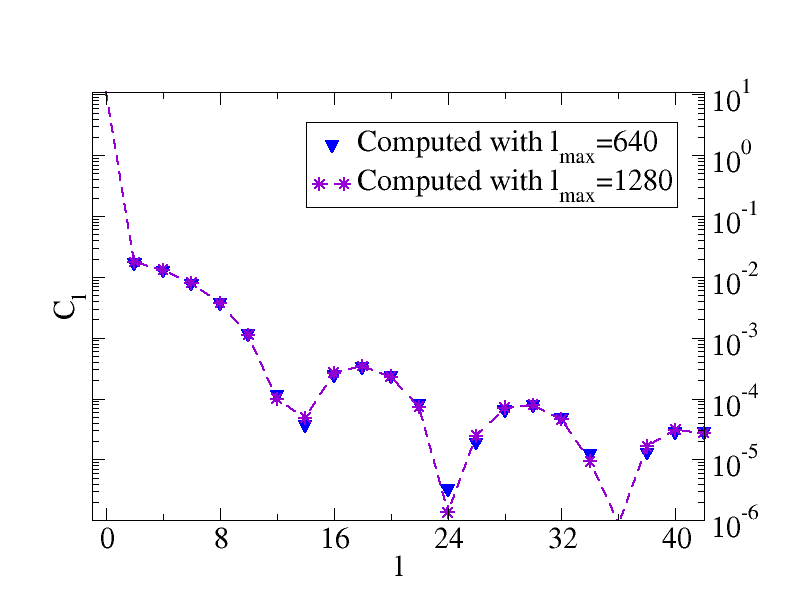}
\caption{\label{fig:ITSacceptance} Angular spectrum of the ITS rapidity acceptance function (the analogous of Eq.~[\ref{acepteq}] that substitutes $45^o\to 73.8^o$). Only even $l$s are shown as the odd $C_l$s vanish by reflection symmetry about the equatorial plane. In comparing with figure~\ref{onlylatcut} (spectrum of the TPC acceptance) we see that all but $C_0$ are now smaller due to the larger rapidity coverage.}
\end{figure}
Since the acceptance of the ITS is broader than that of the TPC (which is slightly less than a unit of rapidity from the equator towards each beam), we expect the equivalent of Eq.~(\ref{acepteq})
with the change $45^o\to 73.8^o$, closer to spherical symmetry, to yield a larger $C_0$ and smaller $C_{l>0}$. Also, since $180/(90-73.8) \sim 10-12$, we expect that the even $C_l$ present some sort of wavy behavior with a period of that order. These features are clearly visible in figure \ref{fig:ITSacceptance}.

The effect of the acceptance is usually very significant. It would be interesting to cut the sky map and see how the CMB power spectrum would change; we imagine that the high multipoles would not, whereas the lower ones would be significantly affected~\footnote{We thank Prof. Naselsky for this comment.}. This can be seen from the drop with $l$ in figure~\ref{latandbars}. We are trying to carry out this exercise for future reporting.

\begin{figure}[hbtp]
\centering
\includegraphics[width=\columnwidth]{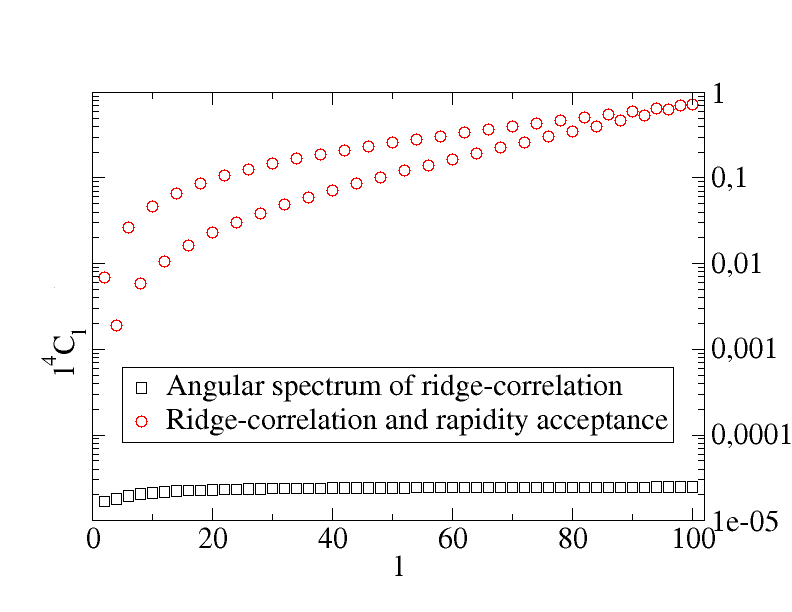}
\caption{ Angular spectrum of a simple function~\cite{Milano:2014qua} mocking the known rapidity ``ridge'' $f(\eta,\phi)=0.033+0.001\cos(2\phi)$ (bottom, squares) and the same function multiplied by the nominal TPC rapidity acceptance cut around 45 degrees latitude (top, circumferences). The effect of the rapidity cut is seen to be overwhelming. 
}
\label{ridgeangspct}
\end{figure}

 As a closer example in RHIC-E, let us plot the 
angular spectrum from a simple function~\cite{Milano:2014qua} that reproduces the ridge correlation, 
$f(\eta,\phi)=0.033+0.001\cos(2\phi)$.
Odd values of $l$ vanish and are not shown, nor is $l=0$ that, though finite and large, vanishes upon representing $l^4 C_l$. This combination is useful to see that the Spherical Harmonic Transform of the ridge (squares in the lower part of figure~\ref{ridgeangspct}) follow, for large $l$, a power law
$C_l\propto l^3$.

However, upon imposing the rapidity cut of ALICE's TPC, which we effect by multiplying  $f$ by a step function  $\theta(\eta_0-|\eta|)$ with $\eta\simeq 1$, the angular spectrum changes very much (circles in the upper part of the plot).

One would hope to be able to disentangle the effect of the acceptance function $A(\theta,\phi)$ (relatively well known) from that of the actual data to describe the $F(\theta,\phi)$ distribution which contains the physical knowledge about the collision. This is however a mathematically ill-posed problem. 
 
The reason is that the expansion coefficients are convoluted upon transforming,
\begin{equation}\label{almFA}
\begin{split}
a_{lm}=&\int d\Omega\ Y_{lm}^*(\theta,\phi)F(\theta,\phi)A(\theta,\phi)\\
=&\int d\Omega\ \sum_{l_F,m_F}\sum_{l_A,m_A}\\
&Y_{lm}^*(\theta,\phi)f_{l_F,m_F}Y_{l_F,m_F}(\theta,\phi)A_{l_A,m_A}Y_{l_A,m_A}(\theta,\phi)\\
=&\sum_{l_F,m_F}\sum_{l_A,m_A} f_{l_F,m_F}A_{l_A,m_A}\left\langle lm|l_Fm_Fl_Am_A\right\rangle \ .
\end{split}
\end{equation}\\
Note the Clebsch-Gordan coefficient in the last line that links any $l$s satisfying the triangle
inequality. The exception is obviously the monopole term, 
for $l_A=m_A=0\longrightarrow\left\langle lm|l_Fm_Fl_Am_A\right\rangle=\delta_{l,l_F}\delta_{m,m_F}$,  $a_{lm}=f_{lm}A_{00}$; in this case alone\footnote{Naselsky {\it et al.}~\cite{Naselsky:2012nw} attempt a separation of a flow part and a statistically fluctuating part to obtain the flow coefficients $v_n$ that is afflicted by this problem, as briefly discussed in section~\ref{sec:interpret}.} an acceptance correction can be applied
$\tilde{a}_{lm}=a_{lm}/A_{00}$.

\section{$P_t$ distributions over the sphere}
Monte Carlo simulations such as HIJING are known to be lacking many features of actual data, such as realistic flow (that should presumably appear in low-harmonics) or otherwise collective effects.

\begin{figure}[hbtp]
\centering
\includegraphics[width=0.9\columnwidth]{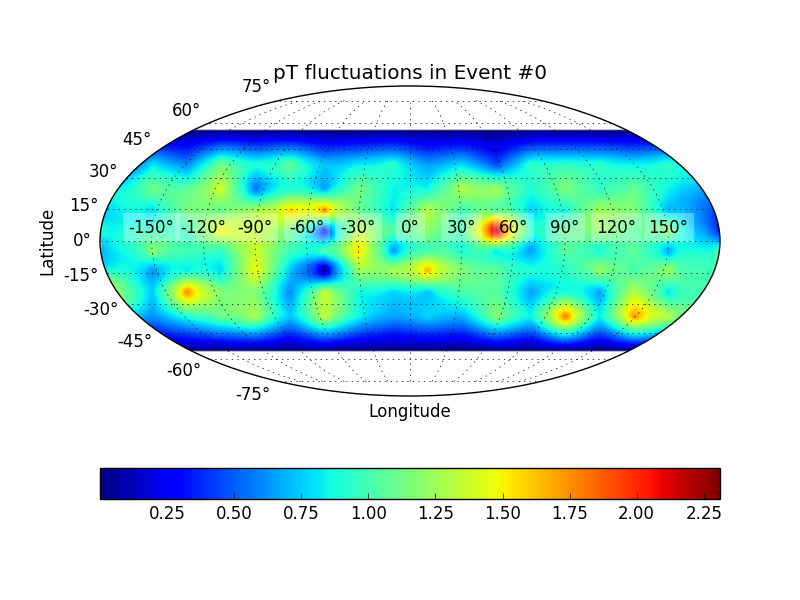}
\includegraphics[width=0.9\columnwidth]{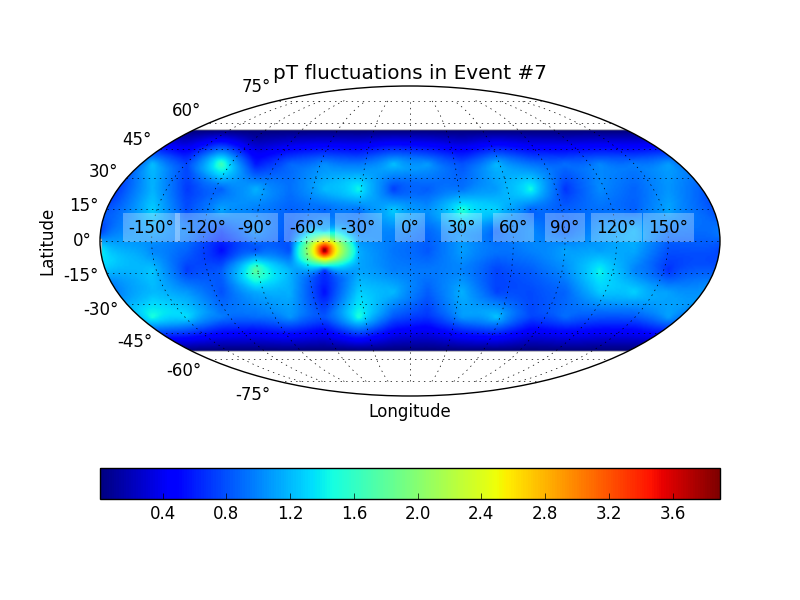}
\includegraphics[width=0.9\columnwidth]{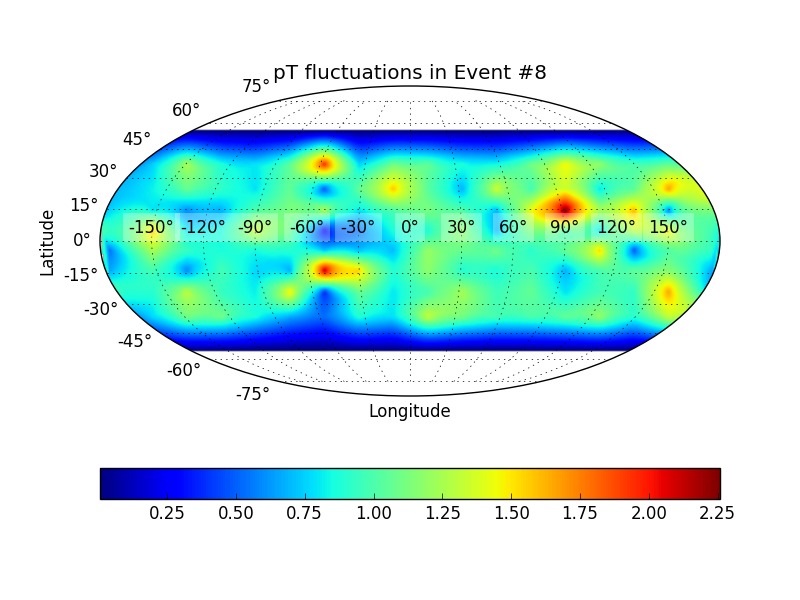}
\caption{Map of $p_t$ for three events (0, 7 and 8 of 186) in Mollweide projection.
}
\label{fig:singleevents}
\end{figure}

We then proceed to show actual data distributions which, as far as we know, is done here for the first time.
Here we divide the $(\theta,\phi)$ coordinate domain into patches, average the $p_t$ of all particles found inside each of them, and then plot it in Mollweide projection.
The result is shown in figure~\ref{fig:singleevents}. There, we plot three high-multiplicity ($O(10^3)$ particles) events taken from~\cite{cernopendata}, showing the distribution of their $p_t$ fluctuations over the sphere.

In the figure, some particularly ``hot'' (red) spots identify angular directions in which relatively high-$p_t$ particles (sometimes identifiable as jets) exit the collision.

\begin{figure}[hbtp]
\centering
\includegraphics[width=\columnwidth]{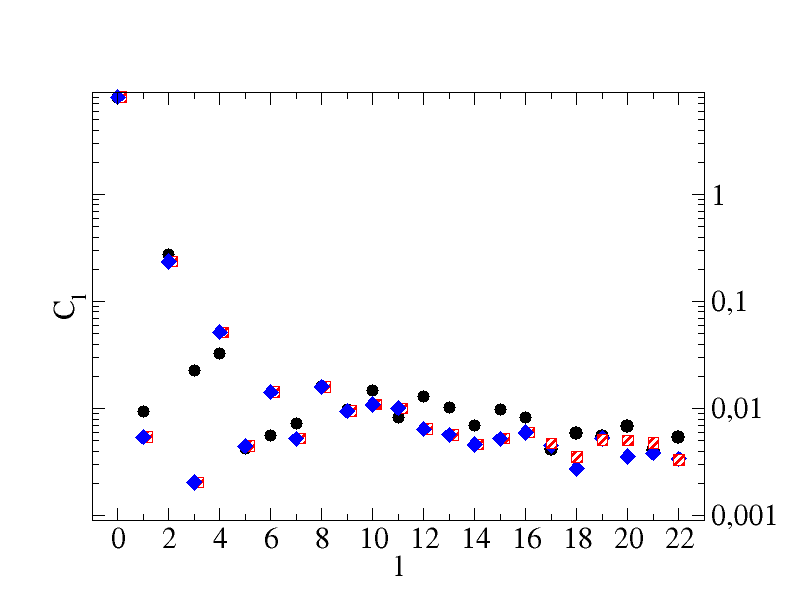}
\caption{ Angular spectrum corresponding to three publically released events (0,7 and 8) taken from~\cite{cernopendata}, for the autocorrelation of the function $(p_t-\bar{p}_t)/\bar{p}_t$.}
\label{3alevents}
\end{figure}

Additionally, we compute the associated $C_l$ angular spectrum of these three events and plot it 
in figure~\ref{3alevents}. 
To prepare the spectrum, we have subtracted the average $p_t$ and divided by it, obtaining a dimensionless function $\left( p_t(\theta,\phi)/\langle p_t\rangle -1 \right)$. 
With this normalization, the monopole  ($l=0$) is controlled by the acceptance alone. 
If it was due to the rapidity cut alone, its value would be  $2\sqrt{2}\pi$. We find it to be slightly smaller, due to the azimuthal inefficiencies. The quadrupole is also in line with expectations based on acceptance.

What is different is the appreciable dipole $l=1$ contribution. It is known~\cite{Naselsky:2012nw} that reflection symmetry respect to the equatorial plane (mid-rapidity) $(\theta-\pi/2)\to -(\theta-\pi/2)$ leads to vanishing odd multipoles.
Therefore, a nonvanishing dipole may be due to an intrinsic asymmetry in the ALICE reconstruction or to actual physical events (such as back to back jets). With few events it is difficult to conclude either, but we are more inclined towards the jet hypothesis.

For larger multipoles there does not seem to be much of a distinction among even and odd $l$, all being of the same order of magnitude~\footnote{This would be very nice if it reflected initial state fluctuations, known e.g. to generate triangular flow $v_3$.}; we do not distinguish any jump at multiples of $l=18$ due to the bars. This means that ALICE is reconstructing tracks in the supposed ``dead angles'' better than the nominal acceptance of the TPC, maybe thanks to the ITS.

We also observe that the three events provide a very similar spectrum, but not identical: the logarithmic scale hides a factor of order 2 between each event's multipoles.

\subsection{Combining all 186 events} \label{subsec:combining}

Each of the realeased high multiplicity individual ALICE events at 2.7 TeV contains $O(10^3)$ particles (short of the $O(10^4)$ hoped for when the experiment was designed~\cite{Naselsky:2012nw}).
Since the number of spherical harmonics up to $l_{\rm max}$ is $(l+1)^2$, to obtain $C_l$ 
for $l\geq 20$, requires $O(400)$  subdivisions of the sphere (so that the spherical harmonic transform is well defined). However, one wishes to have $O(10)$ particles on average on each of those patches, for the local $p_t$ average may be estimated, and this exhausts the number of particles in an event. 

That means that reaching higher $l$ requires combining several events to have larger multiplicity

As already noted, we have 186 events at our disposal, and the total number of charged particle tracks reported (whose momentum is then distributed among each of the sphere patches) is 87623.
This suggests that we can reach at most $l=O(80-90)$. But there is no obstacle for the ALICE
collaboration to deploy the massive statistics of the LHC and reach much higher $l$, separating also centrality classes and increasing the number of particles to average over in each element of area on the sphere.

\begin{figure}[hbtp]
\centering
\includegraphics[width=\columnwidth]{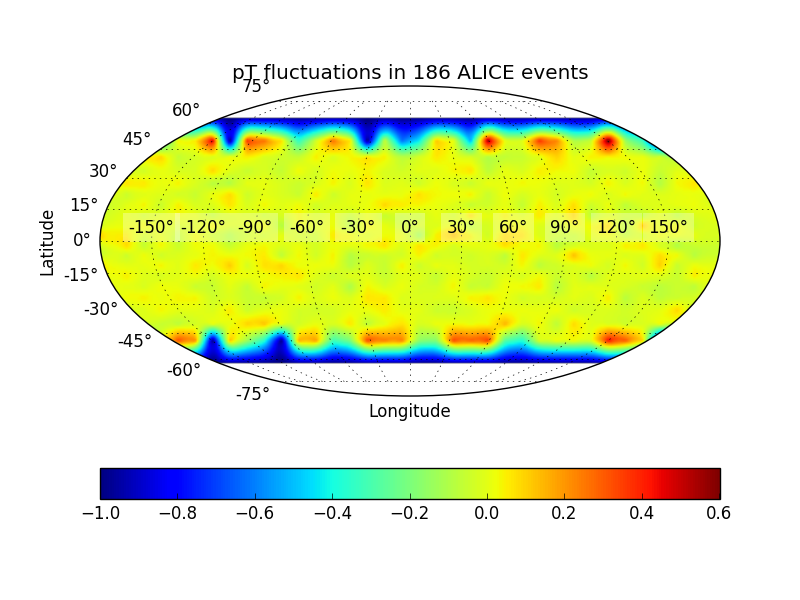}
\includegraphics[width=\columnwidth]{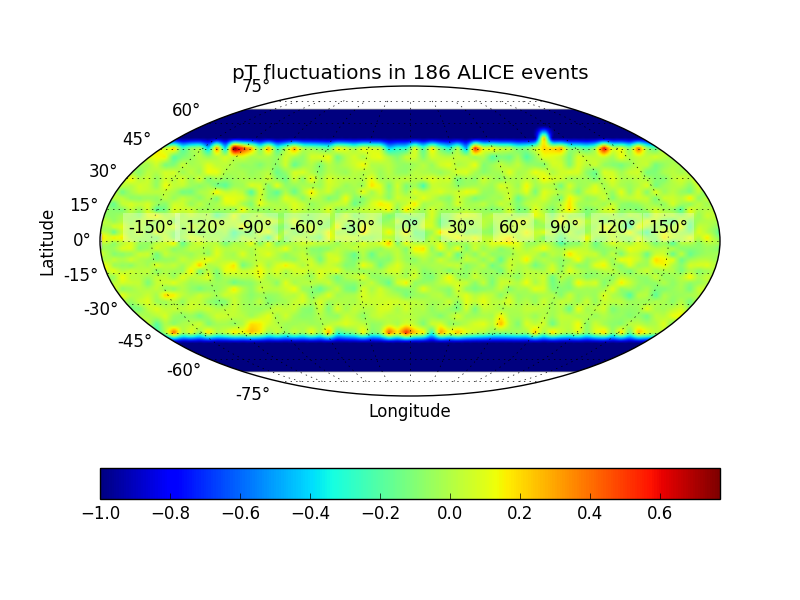}
\includegraphics[width=\columnwidth]{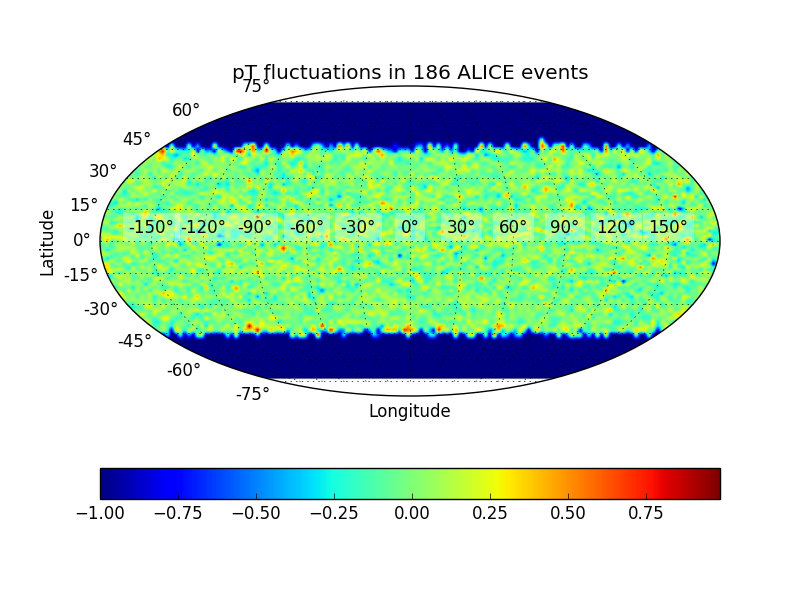}
\caption{Map of $p_t$ for all 186 ALICE events (excepting particles with $p_t>1$ GeV which we have cut off) combined in Mollweide projection, at increasing level of resolution. The number of $(\cos(\theta),\phi)$ subdivisions is, from top to bottom,
$(20,36)$, $(36,72)$, $(72,144)$.
That the high latitudes are blue (low $p_t$) confirms our suspicion that many of the particles there have been detected by the ITS, which measures smaller momenta (those particles exceeding the cut of Eq.~(\ref{acepteq}) are outside the TPC coverage).
}
\label{fig:combinedevents}
\end{figure}

Figure~\ref{fig:combinedevents} displays the $p_t$ distribution of all combined events in Mollweide projection (in analogy to figure~\ref{fig:singleevents} for each separate event), at increasing resolution (smaller $\theta-\phi$ boxes covering the sphere).
Now, the reaction plane of each event forms a random angle with the laboratory reference plane that we used up to this point, so in the absence of a reorientation of each event, the outcome should be, for large number of events, an azimuthally symmetric distribution. To avoid this accident we need to orient the events according to some prefered system of axis associated to each collision and not to the laboratory. One can naturally continue employing the colliding beam axis as $OZ$. 
Then, to choose the $OX$ and $OY$ axis in the perpendicular plane, we use an event by event vector~\cite{Schorner} $Q$ defined by 
\begin{equation}
(Q_x,Q_y)=\sum_{{\rm particles}\ j=1}^{N\ \rm in\ event} {p_t}_{j}\left(\cos (2 \phi_j), \sin (2\phi_j) \right) \label{Qvector} \ .
\end{equation}
Then this vector is oriented by an angle $\Phi\equiv \arctan \left( Q_y/Q_x \right)$ respect to the laboratory $OX$ axis. It is then a simple matter to rotate each event so that all the $Q_x$ axes match, by adding this angle to each particle's azimuth in that event, $\phi_j = \phi_j^{\rm LAB} -\Phi$.
This new angle $\phi_j$ is the one assigned to each particle for the plots in figure~\ref{fig:combinedevents}.

\begin{figure}[hbtp]
\includegraphics[width=\columnwidth]{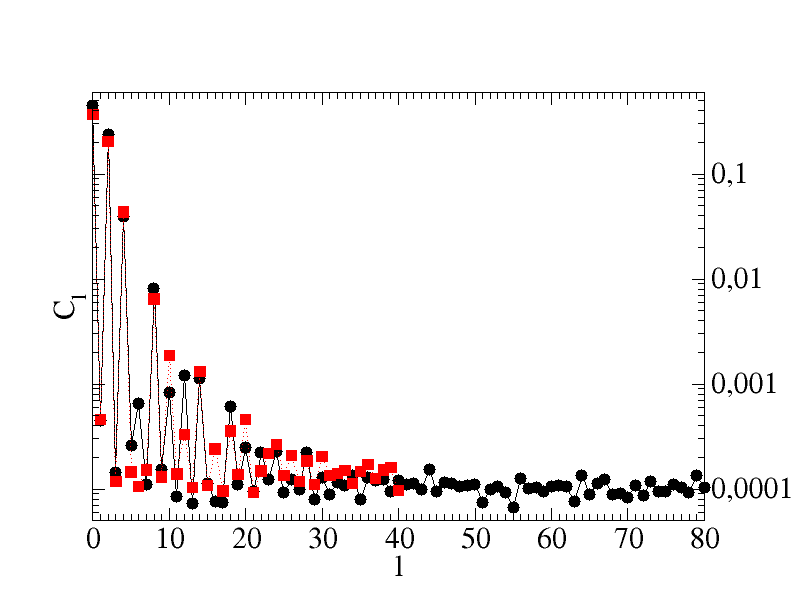}
\includegraphics[width=\columnwidth]{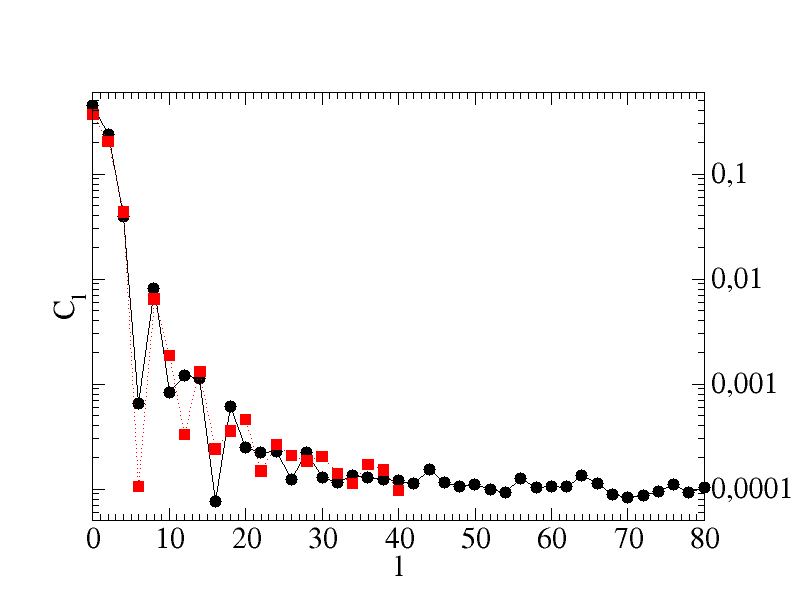}
\caption{\label{fig:final} Angular spectrum of the square fluctuations of the normalized transverse momentum $(p_t-\langle p_t \rangle)/\langle p_t\rangle$ for 186 events made public by the ALICE collaboration. Squares (red online):  $l_{\rm max}=40$. Circles (black online): $l_{\rm max}=80$. The top plot shows both even and odd $l$s, the bottom one only even $l$ for clearer visibility. We display together two computations over the same data but with different $l_{\rm max}$ values, 40 and 80 respectively (as visible).}
\end{figure}

Figure~\ref{fig:final} then shows the angular ``power'' spectrum of the 186 publically released ALICE events, computed once more with SHTOOLS after all events have been quadrupole-aligned. 

We see once more that the odd multipoles are much smaller than the even ones, and that the even ones seem to fall as a power law (so we focus now on the bottom plot, keeping only even $l$), with no acoustic peak distinguishable from noise.

Between $l=36$ and $l=76$ the eye discerns small waves with an alternating 8-12 pattern that are reminiscent of figure~\ref{allbars} due to azimuthal gaps in acceptance; we believe that they are actually caused by the rapidity acceptance cut of the ITS, as argued in section~\ref{sec:acceptance} (it is hard to subscribe azimuthal acceptance as the origin of those soft oscillations once the events have been aligned according to their own quadrupole, as the instrumental effects should be averaged out).

A feature of the figure that remains to be understood is why the intensity of the $l=6$ harmonic is so suppressed.  Because $l=6$ is actually a maximum in figure~\ref{onlylatcut}, and is inconspicuous in figure~\ref{fig:ITSacceptance}, we do not see how this dip might be caused by the acceptance. Other possible causes would be the underlying physics   (that the dip at $l=6$ is heralding a peak at $l=8-10$ seems enticing but unlikely: there is no structure there in the estimate of Mocsy and Sorensen reproduced in figure~\ref{anpwspct}) or simply poor statistics.
It would be interesting to know whether it survives an analysis with a significantly larger number of events.

To conclude the analysis, we try to separate the features of the angular spectrum in figure~\ref{fig:final} that are purely due to the rapidity cut. This is achieved~\cite{Naselsky:2012nw}
by subtracting from each $C_l$ the contribution due to the $a_{l0}$ coefficients in Eq.~(\ref{alminv}), namely $C_l^{m=0}=|a_{l0}|^2/(2l+1)$. Actually, the original motivation of this subtraction in~\cite{Naselsky:2012nw}
was that the Monte Carlo was producing a large excess of particles at high latitudes (near the beam axis) while our problem is the ALICE acceptance cuts that actually zero them out. Either way, the rapidity acceptance is a large effect: we see in figure~\ref{fig:final} that the $C_l$ stabilize at $O(10^{-4})$ only after the quick falloff from the rapidity spectrum in figure~\ref{onlylatcut} takes it below that level, somewhere between $l=20$ and $l=30$.

\begin{figure}[hbtp]
\includegraphics[width=\columnwidth]{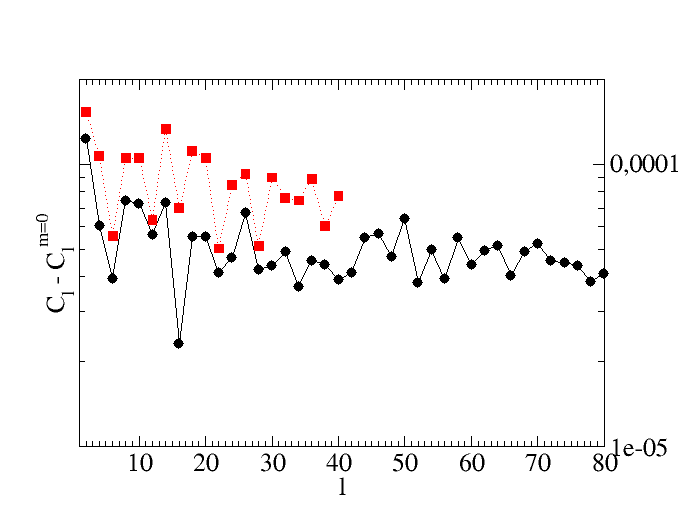}
\caption{\label{fig:spectrumwithoutm0} From the $C_l$ in the lower plot of figure~\ref{fig:final} (with the same symbols), we subtract the contribution due to $m=0$ for each $l$. This takes care of a large part of the rapidity dependence without altering the strength due to the azimuthal dependence.}
\end{figure}

The resulting, subtracted $C_l-C_l^{m=0}$ contains only contributions from the azimuthally-dependent part of the original function. Its spectral power is therefore smaller than the original $C_l$ and not directly comparable with what is common in cosmology, but it may nevertheless be informative. We plot this in figure~\ref{fig:spectrumwithoutm0}.

What we appreciate is that the $m=0$ contributions are at least half of the total $C_l$, even for the largest $l$ that we consider, and are orders of magnitude larger than those with $m\neq 0$ for the smallest $l$s. There is a noticeable difference between $l_{\rm max}=40$ and $80$. The drop in $l=6$, even if reduced, is still there; $C_8\sim C_{10} > C_6$. The oscillations with long period of order 8-12 have disappeared, which supports the guess that they come from the ITS rapidity coverage (azimuth--independent).
Still, the outcome is very noisy and no conclusions can really be reached, calling for larger statistics. There is no clear acoustic peak towering above the data.

From figure~\ref{fig:spectrumwithoutm0} we discern that there may be merit in analyzing data with higher statistics: the effect of the rapidity cut does not obscure the higher $l$ coefficients, the $C_l$s remain at the level of $10^{-4}$ above $l=20-30$ when the effect of the rapidity acceptance has dropped below that level.

\section{Further analysis that may be carried out with full statistics} \label{sec:interpret}

Having taken note of the data features that we see in figure~\ref{fig:final}, 
we now dedicate a section to a brief discussion and overview of some recent literature with focus on what to make of the (for now apparently absent) acoustic peaks if any were found, or how to interpret their continued absence, and to physically motivate further analysis.

In brief, there are various aspects and scenarios worth discussing, that are complementary and sometimes competitive. Perhaps data can bring further insight into the theoretical and Monte Carlo based discussion.

In subsection~\ref{subsec:etas} we echo discussions about the damping of sound and the fall of the function $C_l(l)$ with increasing $l$, with the possibility of accessing the viscosity over entropy density ratio if caveats about the effect of acceptance can be circumvented.

Subsection~\ref{subsec:horizon} then addresses superhorizon fluctuations and the possibility expressed in~\cite{Mishra:2007tw} that a peak may appear in the flow $v_l$ coefficients due to an interplay between power-law behaved initial spatial fluctuations and insufficient time to transfer it to a momentum anisotropy.
These flow coefficients may be assessed with standard means in the field, but we dedicate subsection~\ref{subsec:flow} to discuss a method~\cite{Naselsky:2012nw} to obtain the flow coefficients precisely from harmonic analysis. It will be interesting to see if the results agree with the usual extractions and if so, whether such peak as~\cite{Mishra:2007tw} proposes is manifest.

Finally, in subsection~\ref{subsec:tension}, following old work on fluctuations in RHIC-E, we conclude that the absence of acoustic-like peaks either disfavors the presence of any attractive restoring force in the medium (among others the formation of domains characteristic of first-order phase transitions) or indicate a small speed of sound.

\subsection{Possible extraction of $\eta/s$} \label{subsec:etas}
The damping of sound in a fluid is governed by the dimensionless shear viscosity to entropy density ratio, $\eta/s$.
If the multipole $l$ is anomalously large or small, it is because the size of the perturbation with wavenumber $k$ 
\be \label{ltok}
l = k r_{\rm freeze\ out}
\ee
is correspondingly enhanced or suppressed~\cite{KurkiSuonnio} (in cosmology, because of spacetime curvature, the freeze out radius requires more careful definition).
The Silk damping in figure~\ref{specTTwmap} corresponds precisely to the dissipation of sound energy by viscous processes in the fluid.

The damping of a perturbation (in the momentum-stress tensor) in RHIC-E collisions has been characterized~\cite{Staig:2011wj}  by
\be
\delta T_{\mu\nu} = \delta T_{\mu\nu}(0) \times e^{-k^2/k_v^2} e^{ik(x-tc_s)}\ .
\ee
The characteristic attenuation length due to the viscosity is here 
\be \label{Rvtovisco}
R_v = k_v^{-1} = \sqrt{(2/3) (\eta/s) (\tau_f/T)}\ .
\ee
The typical temperature at freeze out is $T\sim 150 $ MeV; $\eta/s\sim 1/(4\pi)$; and 
the evolution time up to freeze out $\tau_f$ requires a  moment's discussion.

The system size at freeze out, 
$r_{\rm freeze\ out}$, is obtained, for example, from Hanbury-Brown-Twiss interferometry at ALICE~\cite{Graczykowski:2014hoa} and is about 6-8 fm. But this system size includes the increase due to the expansion as well as the initial size of the fireball, so one expects $\tau_f<r_{\rm freeze\ out}$. Given the errors inherent to the HBT method and that lead nuclei are already sizeable, it is reasonable to take $\tau_f\sim 3$ fm. One can also correct for the initial equilibration time $\tau_{\rm eq}$.
Then, Eq.~(\ref{Rvtovisco}) produces about $R_v\sim 0.46$ fm.

We may also  note that the sound horizon, $H_s = \int_{\tau_i}^{\tau_f} d\tau c_s(\tau)$, with the sound speed, though not constant, taken of order 0.1-0.2 of the speed of light (unity), is of the same order or less, $H_s\sim (3{\rm fm}-1{\rm fm})\times (0.15\pm 0.05) =(0.3\pm 0.1){\rm fm}\simeq R_v$. 

That $R_v$ is comparable to $H_s$, by itself, means that one should not expect acoustic peaks too much. Those perturbations with wavelength $(\lambda/(2\pi)) = (1/k)< R_v$ will be damped by viscosity and not be prominent in the spectrum. On the other hand, those with $(1/k)>R_v$ are not yet strongly damped, but the scale is so large that serious damping occurs already for very low multipoles $l\sim 4-6$. (To see it, note that by virtue of Eq.~(\ref{ltok}), $R_v<(\tau_f-\tau_i)/l$, so that $l<(2-3){\rm fm}/0.46{\rm fm}$ ). Also because the undissipated modes simultaneously satisfy $(1/k)>H_s$, they are ``outside the sound horizon'', that is, they have not been able to propagate a significant fraction of their wavelength (see the next subsection~\ref{subsec:horizon}).

\begin{figure}
\includegraphics[width=\columnwidth]{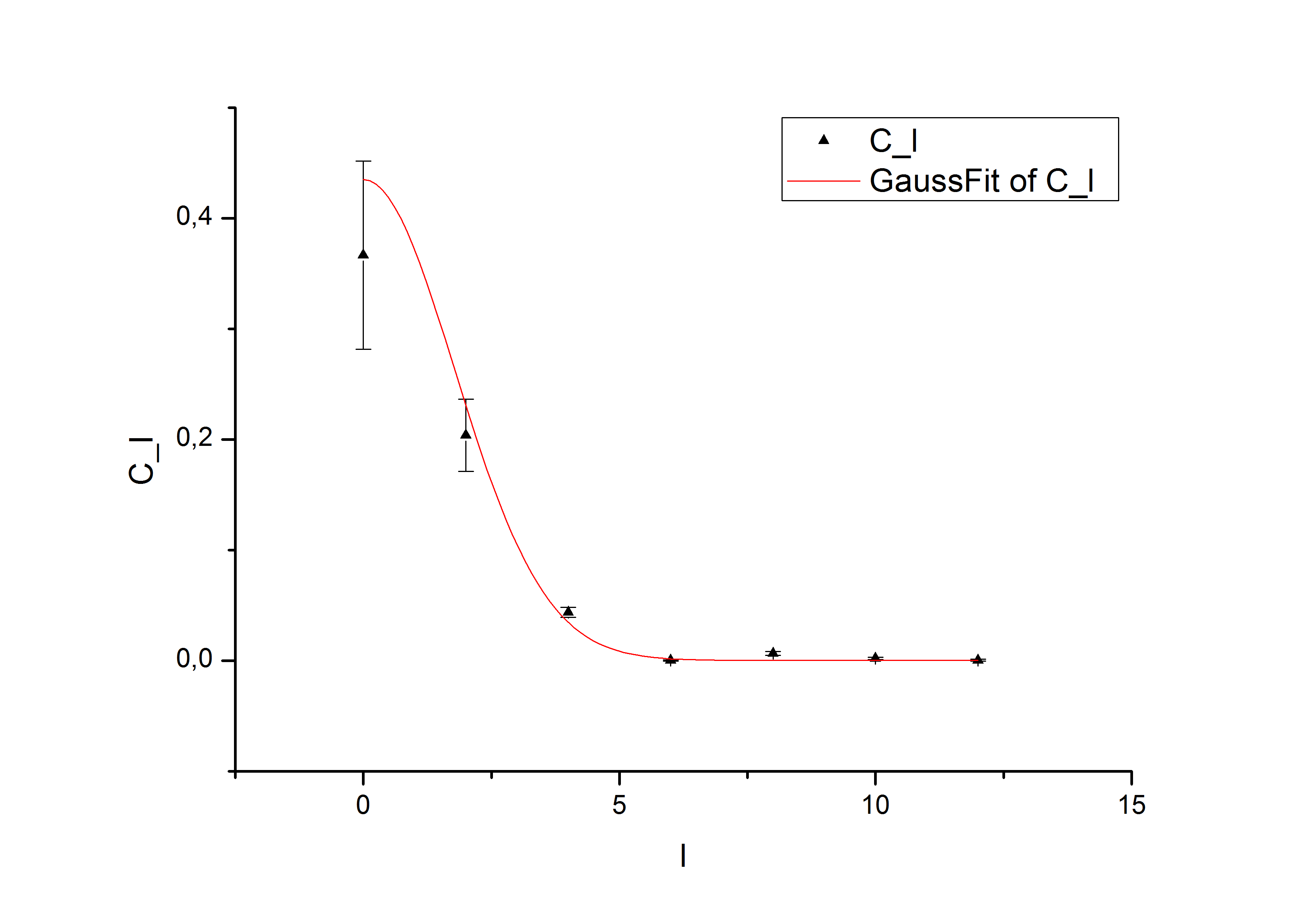}
\caption{Gaussian fit to the data in the lower plot of figure~\ref{fig:final} (basically the contributing $l$s are the even ones up to $l=10$). \label{fig:gaussianfit}}
\end{figure}

If, notwithstanding the large rapidity corrections that we have discussed,
we were to close our eyes to them and take the data in figure~\ref{fig:final} as physically significant and perform a Gaussian fit $C_l \propto \exp(-k^2/k_v^2)$ or, by employing Eq.~(\ref{ltok}), $C_l\propto \exp(-l^2/l_v^2)$, we could use Eq.~(\ref{Rvtovisco}) in reverse and try to extract $\eta/s$.
What figure~\ref{fig:gaussianfit} reports is just such an exercise, where we have taken the smallest $l_{\rm max}$ as the data and the difference with the largest $l_{\rm max}$ as the error.

Figure~\ref{fig:gaussianfit} presents a Gaussian fit of the form ${\rm const.}\times exp(-(l/l_v)^2)$, with $l_v$ the characteristic angular momentum at which the viscosity has damped the sound wave.

The outcome of the fit is $l_v= (3--4)/\sqrt{2}$ that we immediately translate, thanks to
\be
\frac{\eta}{s} = \frac{3T}{2 \tau_f} \left(\frac{r_{\text{freeze out}}}{l_v}\right)^2\ ,
\ee
and to Eq.~(\ref{ltok}),~(\ref{Rvtovisco}), 
\be \label{visco}
\frac{\eta}{s} \simeq 2.1 \pm 0.8\ .
\ee

This number, with a large uncertainty, is not huge, though 
other extractions of this quantity also based on statistical methods~\cite{Gavin:2007zz} yield results about a factor 10 smaller. Of course we do not take the result of Eq.~(\ref{visco}) seriously in view of the systematic shift due to the rapidity acceptance, and leave it for now as a demonstration of the method rather than as a physical feature. 
It would perhaps be better to extract the viscosity from a sort of an $m=0$ subtracted graph such as~\ref{fig:spectrumwithoutm0}. But with the number of events at our disposal, that data is very noisy, and only the three lowest-$l$ points clearly have $C_l$ falling with $l$, so that a Gaussian fit is dubious. 

\subsection{Horizon entering of sound waves and suppression of superhorizon fluctuations}
\label{subsec:horizon} 

Following the CMB reasoning, the first acoustic oscillation that would have time to swing once before freeze out would appear as a characteristic acoustic peak in the angular spectrum.
Its wavelength would be comparable to the sound travel distance $H_s=c_s(\tau_f-r_{\rm eq}/c)$ defined in subsection~\ref{subsec:etas}.
Thus, the position of the first acoustic peak, if one was found, at $l_{\rm min}$, could be related to the speed of sound by
\be \label{speedsound}
\frac{c_s}{c} \sim \frac{2\pi}{l_{\rm min}}\frac{r_{\rm freeze\ out}}{c\tau_f-r^{\rm equilibration}}\ .
\ee
Taking a typical freeze out radius of 6 fm~\cite{Graczykowski:2014hoa}, an equilibration radius of 1 fm, and a speed of sound $c_s\sim (0.1-0.2)c$, this would put $l_{\rm min}\sim O(50-100))$ (in the unrealistic case that $c_s\sim c$, then $l_{\rm min}\sim 5-20$).
The amplitude of this presumed peak is proportional to that of any initial fluctuations. But its position in $l$ combines information about the time ellapsed and the sound horizon at freeze out (hence the sound speed).
Eq.~(\ref{speedsound}) would be a way of obtaining the sound speed if an acoustic peak was found in the $C_l$ spectrum of RHIC-E.

While we have focused on $p_t$, the angular analysis methods can be applied to other observables, and the reasoning can be carried over. For example, U. Heinz~\cite{Heinz:2013wva} has compared flow coefficients in lead-lead collisions
with the angular spectra of the CMB. Below, in subsection~\ref{subsec:flow}, we quickly review the method of Naselsky {\it et al.} to extract those flow coefficients. 
In the plots of Heinz, no acoustic peak is seen either.

In very central collisions where all anisotropy comes from fluctuations, it is sometimes proposed~\cite{Mishra:2007tw} that the small-$l$ flow coefficients appear as if they were damped.
 The reason is the lack of time to transfer anisotropy from the spatial to the momentum degree of freedom~\cite{Mishra:2007tw} for modes of large wavelength. Those authors reason that if in the initial conditions a certain perturbation characterized by $l$ has a maximum of spatial anisotropy, it will start decreasing just by its natural oscillation, while the corresponding $l$ momentum anisotropy coefficient will start growing. If no time is given for the amplitude to swing again (with opposite sign) to its full value, the momentum anisotropy will appear attenuated respect to what it could have achieved.
This happens for modes of wavelength large compared with the acoustic horizon 
$H_s = c_s (\tau_{\rm freeze\ out}-\tau_{\rm equilibration})\ll \lambda/2$. The amplitude of the perturbation will be suppressed by a factor $2H_s/\lambda$,
so that 
\be
\langle v_l\rangle_{\rm observed} = {\rm constant}\times l \times \langle v_l\rangle_{\rm initial \ maximum}\ .
\ee
This predicted increase of $v_l$ with $l$ is something that an analysis of ALICE data could check. It is of note that hydrodynamic arguments (the very concept of flow) must break when the wavelength is too small and one resolves individual particles. From Eq.~(\ref{ltok}), and assuming that one should not speak of a fluid for lengths below 1 fm, $l<l_{\rm fluid} = 2\pi r_{\rm freeze \ out}/\lambda\sim 30$ it is clear that any coefficients $C_l$ or $v_l$ above this number possibly do not have fluid properties and must come from statistical fluctuations.

Saumia and Srivastava~\cite{Saumia:2015lda} have reported Monte Carlo computations where initial fluctuations follow a power law falling with $l$, but the resulting momentum anisotropies $v_l$ do not develop for small $l$ and thus a maximum appears for a small $l$ of order 4-6. 
Their calculation is in the laboratory fixed frame; the difference is that we have
taken into account of the strong effect of the acceptance that as shown in section~\ref{sec:acceptance}. Still, since they are concentrating on an azimuthal anisotropy~\cite{Mishra:2008dm} and not on rapidity, an analysis by the ALICE collaboration employing tracks without demanding a hit in the innermost layer of the ITS can avoid most of the acceptance effect and test the scenario. 
It remains to see how to extract enough flow coefficients for an analysis, but if this turns out to be correct, the position of the peak will allow to read off the power-law strength of the initial fluctuations.

Now, Sorensen~\cite{Sorensen:2008dm} has argued that this behavior of the flow coefficients $v_l$ could also be visible in the $C_l(p_t)$ angular spectrum of the $p_t$ distribution. He refers to this as a ``valley'' in the $C_l$ {\it versus} $l$ graph. 
This is not visible in our figure~\ref{fig:final}. The first three $C_l$ values \emph{decrease with increasing $l$}, all the opposite of the behavior suggested by the superhorizon argument. This would suggest that what we see right now in figure~\ref{fig:final} might be something different from collective flow, and it would be interesting for further ALICE measurements to confirm it.

\subsection{Extraction of single-event flow coefficients}\label{subsec:flow}

The authors of~\cite{Naselsky:2012nw} provide a method of extracting the flow coefficients 
from the spherical harmonic expansion in Eq.~(\ref{alminv}) but for the \emph{particle number} $N$ instead of the $p_t$ distribution that we have been pursuing.  Though we have not calculated this directly yet, we mention it very briefly as it is a closely related observable and we may address it in the near future with the reduced public statistics, though we hope that the experimental ALICE collaboration will do it better. It requires high multiplicity events ($N> v_n^{-2}$), so that small $v_n$'s may not be reachable.

The principle is an \emph{ad hoc} factorization
\be
\frac{d^2 N}{d\phi d\eta} = F(\theta,\phi) \left[ 1+ 2\sum_n v_n \cos (n(\phi-\Psi_n)) \right] 
\ee
where the right hand side is divided into a ``non-flow'' stochastic piece $F$ and a ``flow'' part in the bracket. (The $\Psi_n$ serves the purpose of orienting the event plane, providing a reference to determine the $\phi$ angle in the collision frame; for $n=2$ it is equivalent to the $Q$-vector orientation that we have employed above in subsection~\ref{subsec:combining}.)

If the harmonic decomposition of the functions $ \frac{d^2 N}{d\phi d\eta}$ and $F$ is
given by the coefficients $\nu_{l,m}$ and $f_{l,m}$ respectively, 
those authors deduce\footnote{Here, 
$g_n=2\pi N_{n,0}N_{n,n} \int_{-1}^1 dx P_n^0(x) P_n^n(x)$  with the standard conventions for complex spherical harmonics,\\
\[N_{n,0}=\sqrt{\frac{(2n+1)}{4\pi}}\ ,\ \ \ N_{n,n}=\sqrt{\frac{(2n+1)}{4\pi(2n)!}}\ .\]
(We remark that SHTOOLS uses real spherical harmonics, so the normalization differs).\\
}
\begin{equation}
\nu_{n,n} = f_{n,n} + v_n \nu_{n,0} g(n) e^{in\Psi_n}\ .
\end{equation}
From this expression, \emph{if the flow signal is large enough} so that $|f_{n,n}|$, the stochastic part, is negligible, they conclude that 
\begin{equation}
v_n \simeq \frac{|\nu_{n,n}|}{g(n) |\nu_{n,0}|}
\end{equation} 
This is practical because only the coefficients of the data expansion, $\nu_{l,m}$ appear; the $F$ function is not directly observable, as one cannot hope to recognize the flow and nonflow parts of the momentum distribution in a single event. As the Monte Carlo generators do not handle flow well, $F$ may perhaps be directly characterized from the HIJING particle distribution, but it is best avoided altogether.

\subsection{Absence of acoustic peaks at low $l$: compatible with a crossover transition from QGP to a hadron gas} \label{subsec:tension}

In early proposals~\cite{Mishra:2007tw} to study the $C_l$ angular power spectrum in RHIC-E, 
the correspondence with superhorizon fluctuations in cosmology was noted. 
Their use of only very central collisions means that those authors defaulted to using a lab-fixed frame and not a collision-intrinsic frame.  As we have seen, efforts in that direction face 
important obstacles from the acceptance function. 

Those authors also noted that at the large scales probed by the elliptic flow there are no acoustic oscillations, but wondered whether they could appear at smaller scales. They proposed
that a restoring force could be provided by the surface tension at the wall of bubbles between the QGP and the hadron gas \emph{if the phase transition was first order.}
This was tenable in the early 90s. For example, there were reported computations of the surface tension in lattice Quantum Chromodynamics~\cite{Huang:1990jf}
by studying the nucleation of the presumed bubbles of the hadron gas in the QGP.
The surface tension $\gamma$ appears in Laplace's law for a spherical bubble, $p_{\rm in}-p_{\rm out} = 2\gamma/r$. The computations for $N_f=2$ and for equal QGP and hadron temperature gave~\cite{Huang:1990jf} $\gamma = 0.277(88)T_c^3$ in terms of the critical transition temperature $T_c$.
From here, a ballpark value of $50$ MeV/fm$^3$ was used in~\cite{Digal:1997fx} for $T_c\sim 170$ MeV.

All this theory is less supported today in ALICE conditions, as lattice evidence has built up that the transition in RHIC-E at low baryon density is a crossover. But today the argument can perhaps be reversed, and the absence of acoustic peaks be used to constrain the value of the surface tension $\gamma$ that we presume to be zero, but it would be interesting to bind its value from above. 
Additionally, the RHIC Beam Energy Scan~\cite{Odyniec:2015iaa}, at lower energies and higher baryon densities, might cross the line of the first order phase transition  beyond the critical endpoint in the QCD phase diagram.
Thus, let us spend a few lines relating the position of the peak to the presumed $\gamma$.

Hadron gas bubbles of a first order phase transition would nucleate in the QGP at the transition temperature $T_c$.
One such bubble of radius $r$ would be affected by forces due to the pressure difference between the outside and the inside of the bubble,
$\Delta P = (P_{\rm QGP}-P_{\rm Hadron})$
and from the surface tension,
\be
F_r^\gamma -F_r^p = -4\pi r^2 \Delta P  -8\pi r\gamma \ .
\ee
From this, the bubble would have a critical stability radius at vanishing net force given by
\be
r_{\rm eq} = \frac{-2\gamma}{\Delta P}\ .
\ee
If the bubble was compressed by a perturbation, $r=r_{\rm eq} + r_{\rm pert}$, the bubble would oscillate according to
\be
(\rho +P)_{\rm QGP} \frac{d^2r_{\rm pert}}{dt^2} = -8\pi (\gamma + r_{\rm eq}\Delta P)r_{\rm pert}\ ,
\ee
that is, with angular frequency
\be
\omega = \sqrt{\frac{-8\pi\gamma}{(\rho+P)_{\rm QGP}}}\ .
\ee
Nonobservation of oscillations up to a certain $\omega$ could then constrain the square root of that surface tension $\gamma$.

Since $k\sim-8\pi\gamma$, eq. (\ref{ltok}) tells us that $l/r_{\text{freeze out}}\sim-8\pi\gamma$; if no peak is visible at small $l$
\be
(-\gamma)<\frac{l_{\text{no peak}}}{8\pi r_{\text{freeze out}}}
\ee
Speculating that the feature at $l=8$ is meaningless, but having it as uncertainly, $-\gamma<8/(8\pi r_{\text{freeze out}})\simeq1/20\text{ fm}^{-1}$.

\subsection{Conclusions}

We believe that we have focused many elements from previous, inspirational works, into a document that can be useful for the ALICE collaboration or other RHIC-E experiments. Particularly, with the small set of released data that we have been able to use, we have tried some of the analysis that we propose for deployment.

Our work is distinct from earlier contributions in several respects.
Unlike~\cite{Mishra:2007tw,Saumia:2015lda}, we have obtained the angular spectrum of $C_l$; we have used actual ALICE data instead of Monte Carlo HIJING simulation; we have performed a brief acceptance analysis for individual events refered to the lab frame and even more briefly for events combined in the reaction frame; and we have employed $p_t$, which is an actual observable in RHIC-E, instead of the temperature $T$ that must always be deduced from other data.
Unlike~\cite{Naselsky:2012nw}, we have performed actual acceptance cuts; studied actual data; and focused on fluctuations rather than trying to separate the flow with the little statistics we have.
However, much remains to be done in upcoming work.

Our results are limited by the small number of publically available events: this hinders us from performing a separation of the events in centrality classes, for example (useful to separate pure fluctuations from anisotropy-induced flow moments).

The ALICE collaboration, with much more statistics, could calculate the dispersion of the angular spectra (the cosmic variance in cosmology is estimated with only one sample).

 We also suggest that the $m=0$  subtraction from~\cite{Naselsky:2012nw} 
would capture most of what we have called the ``rapidity'' cut and allow closer examination of other structures that might have a more physical origin. And we have indeed performed this subtraction and shown the drastic decrease of low-$l$ $C_l$s; for larger $l$ there is no effect.
This procedure does not change the dependence on the azimuthal asymmetry at all, and it was employed in~\cite{Naselsky:2012nw} to isolate flow azimuthal signals injected by hand in the Monte Carlo simulation. 

A further improvement that the ALICE collaboration can carry on is to avoid demanding a hit in the innermost layer of the ITS. Our guess is that this will diminish the polution from the azimuthal acceptance.

Finally, we have exposed what physics may be explored with a full analysis: angular spectra of $p_t$, or of particle number to extract flow coefficients; study of viscosity over entropy density (from damping), or, from the presence/absence of an acoustic peak, the speed of sound or the lack of a first order phase transition.

{\emph{We thank V. Gonz\'alez for bringing the public ALICE data~\cite{cernopendata} to our attention, and both him and P. Ladr\'on de Guevara for useful conversations. We also thank J. Torres-Rinc\'on for feedback on the first draft of this manuscript. Work supported by 
the Spanish networks on Hadronic Physics (FIS2014-57026-REDT) and Consolider Centro Nacional de Fisica de Part\'{\i}culas, astropart\'{\i}culas y nuclear, (CPAN FPA2015-69037-REDC),
and FJLE by spanish grant FPA2011-27853-C02-01}}.

 

\end{document}